\documentclass[12pt]{iopart}

\usepackage{graphicx}
\usepackage{iopams}

\begin{document}
\newcommand{\ve}{\varepsilon}
\newcommand{\ri}{{\rm i}}

\thispagestyle{empty}

\title[Casimir free energy of dielectric films]{Casimir free energy
of dielectric films: Classical limit, low-temperature behavior and control}

\author{G~L~Klimchitskaya${}^{1,2}$ and V~M~Mostepanenko${}^{1,2,3}$
}

\address{${}^1$Central Astronomical Observatory at Pulkovo of the
Russian Academy of Sciences, Saint Petersburg,
196140, Russia}
\address{${}^2$Institute of Physics, Nanotechnology and
Telecommunications, Peter the Great Saint Petersburg
Polytechnic University, Saint Petersburg, 195251, Russia}
\address{${}^3$Kazan Federal University, Kazan, 420008, Russia}

\ead{g.klimchitskaya@gmail.com}

\begin{abstract}
The Casimir free energy of dielectric films, both free-standing in vacuum
and deposited on metallic or dielectric plates, is investigated. It is
shown that the values of the free energy depend considerably on whether
the calculation approach used neglects or takes into account the dc
conductivity of film material. We demonstrate that there are the
material-dependent and universal classical limits in the former and
latter cases, respectively. The analytic behavior of the Casimir
free energy and entropy for a free-standing dielectric film at low
temperature is found. According to our results, the Casimir entropy
goes to zero when the temperature vanishes if the calculation approach
with neglected dc conductivity of a film is employed. If the dc
conductivity is taken into account, the Casimir entropy takes the
positive value at zero temperature, depending on the parameters of a
film, i.e., the Nernst heat theorem is violated. By considering the
Casimir free energy of SiO${}_{2}$ and Al${}_{2}$O${}_{3}$ films deposited
on a Au plate in the framework of two calculation approaches, we
argue that physically correct values are obtained by disregarding
the role of dc conductivity. A comparison with the well known results
for the configuration of two parallel plates is made. Finally, we
compute the Casimir free energy of SiO${}_{2}$, Al${}_{2}$O${}_{3}$ and
Ge films deposited on high-resistivity Si plates of different
thicknesses and demonstrate that it can be positive, negative and
equal to zero. The effect of illumination of a Si plate with laser
light is considered. Possible applications of the obtained results to
thin films used in microelectronics are discussed.
\end{abstract}
\pacs{77.55.-g, 77.22.Ch}
\noindent{\it Keywords:\/} Casimir free energy, dielectric films, Nernst heat
theorem, dc conductivity.
\submitto{\JPCM}

\maketitle

\section{Introduction}

The fluctuation-induced phenomena and, specifically, the van der
Waals and Casimir forces, play a progressively increasing role in
many topics of physics, chemistry and biology (see the monographs
\cite{1,2} and reviews \cite{3,4,5,6}. They are responsible for
interaction of electrically neutral, but polarizable, particles
with material surfaces \cite{7,8,9,10}, find applications in
nanoscience \cite{11,12,13,13a}, play important role in many effects
of condensed matter physics \cite{14,15,16,17,18,19}, and are even
used in elementary particle physics for constraining some theoretical
predictions beyond the standard model \cite{20,21,22}.

When considering the Casimir interaction, the configuration of two
material bodies interacting through a vacuum gap is in most common
use. Thus, the fundamental theory of the van der Waals and Casimir
forces (the Lifshitz theory) was originally formulated for two
parallel semispaces separated by some distance $a$ \cite{1,2,23}.
Later on, this theory was generalized for the cases of arbitrarily
many plane parallel material layers and closely spaced surfaces of
any geometrical shape \cite{2,6}.

Another system described by the Lifshitz theory is a free-standing
or deposited on a thick plate material film. Electromagnetic
fluctuations result in some Casimir free energy of a film, which
should be taken into account in the balance of energies responsible
for its stability. This problem was formulated more than 80 years
ago by Derjaguin and has been investigated using several
approximate approaches in physical chemistry of surfaces \cite{24}.

Recently it was shown that the Casimir free energy of both
nonmagnetic and magnetic metallic films calculated rigorously in
the framework of the Lifshitz theory differ considerably depending
on whether the Drude or the plasma model is used for extrapolation
of the optical data for the complex index of refraction to zero
frequency \cite{25,26,27}. It is well known that there is an
outstanding problem in the Lifshitz theory \cite{2,3,6}. The
point is that the low-frequency response of metals to classical
electromagnetic fields is commonly described by the dissipative
Drude model. However, the predictions of the Lifshitz theory are
excluded by the experimental data of all precise experiments on
measuring the Casimir force between two metallic test bodies if
this model is used to describe the response of metals to a
fluctuating field \cite{28,29,30,31,32,33,34,35}. The same
experiments are in a very good agreement with the Lifshitz theory
if the lossless plasma model is used for extrapolation of the
optical data of metals to zero frequency
\cite{28,29,30,31,32,33,34,35,36}.

It must be emphasized that in all precise experiments on measuring
the absolute values of the Casimir force \cite{28,29,30,31,32,33,34}
(or force gradients) at separations below $1\,\mu$m theoretical
predictions of the Lifshitz theory using the Drude and the plasma
models differ by only a few percent. The difference in theoretical
predictions by up to a factor of two is reached only at separations
of a few micrometers, where the absolute value of the measured force
becomes too small. Large differences in the predictions of both
approaches are reached in the recently proposed \cite{37,38,39}
differential force measurements, but in the experiments of this
kind the measured force signal is also rather low \cite{35}. Unlike
the configuration of two metallic plates separated with a gap,
the absolute values of the Casimir free energy and pressure of
metallic films, as thin as of 50 or 100\,nm thickness, can differ by
the factors of hundreds and even thousands when the calculation
approaches using the Drude and the plasma models are used
\cite{25,26,27}. Thus, the Drude-plasma dilemma becomes extremely
important not only from the theoretical point of view, but for
important technological applications as well.

In this paper, we investigate the Casimir free energy of dielectric films,
either free-standing or deposited on metallic or dielectric plate. It is shown
that, in some analogy to already investigated case of metallic films
\cite{25,26,27}, the Lifshitz theory leads to considerably different results
depending on whether one takes into account the conductivity of film
material at nonzero temperature or omits it in calculations. Similar result
has been obtained previously in calculations of the measured Casimir forces
between two bodies separated by a vacuum gap with one of them made of
dielectric material \cite{40,41,42,43,44}, but the differences in theoretical
predictions were much less than we find here for a dielectric film.
It was shown that the Lifshitz theory is excluded by the measurement data if
the conductivity at a constant current (i.e., the dc conductivity) is taken into
account in computations. If the dc conductivity is omitted in computations,
the Lifshitz theory was found to be in a good agreement with the experimental
data \cite{40,41,42,43,44}. One more experiment of this type was
proposed in \cite{44a}.

 Furthermore, for two plates or for an atom above a plate  it has been shown \cite{45,46,47,48,49}
that the Lifshitz theory with included dc conductivity violates the third law of
thermodynamics (the Nernst heat theorem), whereas the same theory is in perfect
agreement with thermodynamics when the dc conductivity is omitted.
According to the Nernst heat theorem, in thermal equilibrium the entropy of
a physical system must go to the universal constant, which does not depend
on the parameters of a system, when the temperature vanishes \cite{49a,49b}.
It is conventional to put this constant equal to zero. The rigorous proof
of the Nernst heat theorem is given in the framework of quantum statistical
physics \cite{49a,49b}. In so doing, entropy at zero temperature proves to be
proportional to the logarithm of the number of states with the lowest
energy. For the nondegenerate dynamical state of lowest energy the entropy
at zero temperature is equal to zero.
Note that for metals with perfect crystal lattices the Lifshitz theory violates
the Nernst theorem when the Drude model is used and satisfies it if the plasma
model is employed \cite{50,51,52,53,54}.
 For metals with impurities the Nernst heat theorem is satisfied because at very low
temperature the Casimir entropy abruptly jumps to zero starting from the negative
value \cite{54a}. This, however, does not solve the problem because the
crystal with perfect crystal lattice is a truly equilibrium system with a nondegenerate
ground state, and the Nernst heat theorem must be satisfied in this case.

Here, we devote main attention to the consideration of the Casimir free energy
rather than the Casimir pressure of a film. The point is that the film stability
is determined by the sign of the total film free energy to which the Casimir free
energy is one of important contributions. The film stability is reached when the
total free energy is negative. Thus, we investigate in detail when the Casimir
contribution to the total free energy changes its sign (under this condition
the Casimir pressure changes its sign as well).

According to our results, the Casimir free energy of dielectric film reaches the
classical limit irrespective of whether it is described with included or
omitted dc conductivity. In the latter case, the Casimir free energy goes to the
universal limit with increasing film thickness, whereas in the former the limiting
value of the Casimir free energy depends on the film material. Unlike the case
of two dielectric plates separated with a vacuum gap, the classical limit is
already reached for relatively thin films of less than $3\,\mu$m thickness at room
temperature.

We find asymptotic expressions for the Casimir free energy of dielectric films
at low temperature. It is shown that for a film made of perfect dielectric
(i.e., having zero electric conductivity at all temperatures) the Casimir
entropy satisfies the Nernst heat theorem. If the dc conductivity, which
 is inherent to any
dielectric material at nonzero temperature, is included in calculations,
the Casimir entropy of a film takes the positive value at zero temperature.
This value depends on the film parameters, i.e., the Nernst heat theorem is
violated. We also perform numerical computations for the Casimir free energy
of dielectric films, both free standing and deposited on metallic plate, and
show that it takes negative and positive values, respectively.

Finally, the possibilities to control the Casimir free energy of dielectric
films deposited on a dielectric plate are investigated. For this purpose,
the free energies of silica, sapphire and germanium films deposited on silicon
plates of different thickness are computed. The role of illumination of a
silicon plate with laser light is considered. It is shown that the Casimir free
energy of a film can take both negative and positive values, but an illumination
of a plate makes it positive with exception of only very thin Ge films.

The obtained results can be used in numerous applications of novel dielectric
films in optics and advanced microelectronics including better understanding
of physical principles of thin film deposition \cite{55}.

\section{General formalism for a dielectric film deposited on material plate}

In this section, we formulate several results of the Lifshitz theory in the form
convenient for application to thin films. We also find the classical limit
for the free energy of a free-standing film and demonstrate its dependence on
an inclusion of the dc conductivity of dielectric material.

\subsection{Lifshitz-type formula for a film}

All the configurations of our interest can be considered as particular cases of
a system consisting of the dielectric film of thickness $a$ with dielectric
permittivity $\varepsilon_f(\omega)$ deposited on a material plate
of thickness $d$ with dielectric permittivity $\varepsilon_p(\omega)$.
We assume that this system is at temperature $T$ in vacuum in thermal equilibrium
with an environment. Then, the Casimir free energy of a film per unit area is
given by the following Lifshits-type formula \cite{2,23}:
\begin{eqnarray}
&&
{\cal F}(a,T)=\frac{k_BT}{2\pi}\sum_{l=0}^{\infty}{\vphantom{\sum}}^{\prime}
\int_{0}^{\infty}\!\!k_{\bot} dk_{\bot}\sum_{\alpha}
\ln\left[1-
r_{\alpha}^{(f,v)}(\ri\xi_l,k_{\bot})\right.
\nonumber  \\
&&~~~~~~\times\left.
R_{\alpha}^{(f,p)}(\ri\xi_l,k_{\bot})
\, e^{-2ak_f(\ri\xi_l,k_{\bot})}\right].
\label{eq1}
\end{eqnarray}
\noindent
Here, $k_B$ is the Boltzmann constant, $k_{\bot}$ is the magnitude of the
projection of the wave vector on the plane of a film, $\xi_l=2\pi k_BTl/\hbar$
with $l=0,\,1,\,2,\,\ldots$ are the Matsubara frequencies, the prime on the
summation sign  in $l$ multiplies the term with $l=0$ by 1/2, and the sum in
$\alpha$ makes a summation on the transverse magnetic (TM) and transverse
electric (TE) polarizations of the electromagnetic field.

The reflection coefficients on the boundary surface between a film and a vacuum
are defined as
\begin{eqnarray}
&&
r_{\rm TM}^{(f,v)}(\ri\xi_l,k_{\bot})=\frac{k_f(\ri\xi_l,k_{\bot})-\varepsilon_{f,l}
\,q(\ri\xi_l,k_{\bot})}{k_f(\ri\xi_l,k_{\bot})+\varepsilon_{f,l}
\,q(\ri\xi_l,k_{\bot})},
\nonumber \\
&&
r_{\rm TE}^{(f,v)}(\ri\xi_l,k_{\bot})=\frac{k_f(\ri\xi_l,k_{\bot})-
q(\ri\xi_l,k_{\bot})}{k_f(\ri\xi_l,k_{\bot})+q(\ri\xi_l,k_{\bot})},
\label{eq2}
\end{eqnarray}
\noindent
where $\varepsilon_{f,l}=\varepsilon_f(\ri\xi_l)$ and
\begin{eqnarray}
&&
k_f(\ri\xi_l,k_{\bot})=\sqrt{k_{\bot}^2+\varepsilon_{f,l}\frac{\xi_l^2}{c^2}},
\nonumber \\
&&
q(\ri\xi_l,k_{\bot})=\sqrt{k_{\bot}^2+\frac{\xi_l^2}{c^2}}.
\label{eq3}
\end{eqnarray}

The reflection coefficients on the boundary surface between a film and a plate
of finite thickness $d$ are given by
\begin{equation}
R_{\alpha}^{(f,p)}(\ri\xi_l,k_{\bot})=\frac{r_{\alpha}^{(f,s)}(\ri\xi_l,k_{\bot})+
r_{\alpha}^{(p,v)}(\ri\xi_l,k_{\bot})e^{-2dk_p(\ri\xi_l,k_{\bot})}}{1+
r_{\alpha}^{(f,s)}(\ri\xi_l,k_{\bot})
r_{\alpha}^{(p,v)}(\ri\xi_l,k_{\bot})e^{-2dk_p(\ri\xi_l,k_{\bot})}},
\label{eq4}
\end{equation}
where the reflection coefficients  $r_{\rm TM,TE}^{(f,s)}$ on the plane between
a film and a semispace, made of the same material as a plate, take the form
\begin{eqnarray}
&&
r_{\rm TM}^{(f,s)}(\ri\xi_l,k_{\bot})=\frac{\varepsilon_{p,l}
k_f(\ri\xi_l,k_{\bot})-\varepsilon_{f,l}
k_p(\ri\xi_l,k_{\bot})}{\varepsilon_{p,l}k_f(\ri\xi_l,k_{\bot})+\varepsilon_{f,l}
k_p(\ri\xi_l,k_{\bot})},
\nonumber \\
&&
r_{\rm TE}^{(f,s)}(\ri\xi_l,k_{\bot})=\frac{k_f(\ri\xi_l,k_{\bot})-
k_p(\ri\xi_l,k_{\bot})}{k_f(\ri\xi_l,k_{\bot})+k_p(\ri\xi_l,k_{\bot})},
\label{eq5}
\end{eqnarray}
\noindent
and
\begin{equation}
k_p(\ri\xi_l,k_{\bot})=\sqrt{k_{\bot}^2+\varepsilon_{p,l}\frac{\xi_l^2}{c^2}}.
\label{eq6}
\end{equation}
\noindent
The remaining coefficients $r_{\alpha}^{(p,v)}$ in (\ref{eq4}) are given
by (\ref{eq2}) with a replacement of the index $f$ (film) with the index
$p$ (plate).

Thick plates can be considered as semispaces. The Casimir free energy of a dielectric
film deposited
on a thick plate is obtained from (\ref{eq1}) and (\ref{eq4}) by limiting
transition $d\to\infty$. In this case we have
\begin{equation}
R_{\rm TM,TE}^{(f,p)}(\ri\xi_l,k_{\bot})=r_{\rm TM,TE}^{(f,s)}(\ri\xi_l,k_{\bot}).
\label{eq7}
\end{equation}

For a free-standing film in vacuum one has $\varepsilon_{p,l}=1$ and
\begin{equation}
R_{\rm TM,TE}^{(f,p)}(\ri\xi_l,k_{\bot})=r_{\rm TM,TE}^{(f,v)}(\ri\xi_l,k_{\bot}).
\label{eq8}
\end{equation}

Below we make the asymptotic expansions and numerical computations using
(\ref{eq1})--(\ref{eq8}). For this purpose it  is convenient to introduce
the dimensionless variables
\begin{eqnarray}
&&
y=2aq(\ri\xi_l,k_{\bot}),
\label{eq9} \\
&&
\zeta_l=\frac{\xi_l}{\omega_c}=\frac{2a\xi_l}{c}
\equiv\tau l,
\nonumber
\end{eqnarray}
\noindent
where an important parameter $\tau$ is given by
\begin{equation}
\tau=4\pi\frac{k_BTa}{\hbar c}.
\label{eq9a}
\end{equation}

In terms of this variables the Casimir free energy of a film takes the form
\begin{eqnarray}
&&
{\cal F}(a,T)=\frac{k_BT}{8\pi a^2}\sum_{l=0}^{\infty}{\vphantom{\sum}}^{\prime}
\int_{\zeta_l}^{\infty}\!\!y\,dy\sum_{\alpha}
\ln\left[1-
r_{\alpha}^{(f,v)}(\ri\zeta_l,y)\right.
\nonumber  \\
&&~~~~~~\times\left.
R_{\alpha}^{(f,p)}(\ri\zeta_l,y)
\, e^{-\sqrt{y^2+(\varepsilon_{f,l}-1)\zeta_l^2}}\right].
\label{eq10}
\end{eqnarray}
\noindent
Here, the reflection coefficients $r_{\alpha}^{(f,v)}$ are
\begin{eqnarray}
&&
r_{\rm TM}^{(f,v)}(\ri\zeta_l,y)=\frac{\sqrt{y^2+(\varepsilon_{f,l}-1)\zeta_l^2}
-\varepsilon_{f,l}y}{\sqrt{y^2+(\varepsilon_{f,l}-1)\zeta_l^2}
+\varepsilon_{f,l}y},
\nonumber \\
&&
r_{\rm TE}^{(f,v)}(\ri\zeta_l,y)=\frac{\sqrt{y^2+(\varepsilon_{f,l}-1)\zeta_l^2}
-y}{\sqrt{y^2+(\varepsilon_{f,l}-1)\zeta_l^2}+y}.
\label{eq11}
\end{eqnarray}
\noindent
The reflection coefficients $R_{\alpha}^{(f,p)}$ expressed in terms of the variables
(\ref{eq9}) are given by (\ref{eq4}) where $\xi_l$ and $k_{\bot}$ are replaced
with $\zeta_l$ and $y$,
\begin{eqnarray}
&&
r_{\rm TM}^{(f,s)}(\ri\zeta_l,y)=\frac{\varepsilon_{p,l}
\sqrt{y^2+(\varepsilon_{f,l}-1)\zeta_l^2}-\varepsilon_{f,l}
\sqrt{y^2+(\varepsilon_{p,l}-1)\zeta_l^2}}{\varepsilon_{p,l}
\sqrt{y^2+(\varepsilon_{f,l}-1)\zeta_l^2}+\varepsilon_{f,l}
\sqrt{y^2+(\varepsilon_{p,l}-1)\zeta_l^2}},
\nonumber \\
&&
r_{\rm TE}^{(f,s)}(\ri\zeta_l,y)=\frac{\sqrt{y^2+(\varepsilon_{f,l}-1)\zeta_l^2}
-\sqrt{y^2+(\varepsilon_{p,l}-1)\zeta_l^2}}{\sqrt{y^2+(\varepsilon_{f,l}-1)\zeta_l^2}
+\sqrt{y^2+(\varepsilon_{p,l}-1)\zeta_l^2}},
\label{eq12}
\end{eqnarray}
\noindent
the reflection coefficients $r_{\rm TM,TE}^{(p.v)}$ are obtained from (\ref{eq11})
by a replacement of the index $f$ for $p$ and
\begin{equation}
k_p(\ri\zeta_l,y)=\frac{1}{2a}\sqrt{y^2+(\varepsilon_{p,l}-1)\zeta_l^2}.
\label{eq13}
\end{equation}

We consider, first, the Casimir free energy (\ref{eq10}) in the limiting case of
large separations (high temperatures). For the configuration of two parallel
plates separated with a gap, the main contribution to the Casimir free energy does
not depend on $\hbar$ in this case, i.e., becomes classical.

\subsection{Classical limit}

It is well known \cite{2} that at sufficiently high temperatures
\begin{equation}
T\gg T_{\rm eff}=\frac{\hbar c}{2ak_B}
\label{eq14}
\end{equation}
\noindent
the zero-frequency term of the Lifshitz formula (\ref{eq10}) gives the dominant
contribution to the Casimir free energy, whereas contributions of all nonzero
Matsubara frequencies become exponentially small. The condition (\ref{eq14})
can be rewritten as $a\gg\hbar c/(2k_BT)$, i.e., as a restriction on the
separation between two parallel plates or on a film thickness depending on the
meaning of the parameter $a$.

For a free-standing dielectric film in vacuum with finite static dielectric
permittivity $\varepsilon_{f,0}=\varepsilon_l(0)<\infty$ the zero-frequency
contribution to the Casimir free energy (\ref{eq10}) is given by
\begin{equation}
{\cal F}^{(l=0)}(a,T)=\frac{k_BT}{16\pi a^2}\int_{0}^{\infty}\!\!\!
ydy\ln(1-r_{f,0}^2e^{-y}),
\label{eq15}
\end{equation}
\noindent
where, according to (\ref{eq8}) and (\ref{eq11}),
\begin{eqnarray}
&&
r_{\rm TM}^{(f,v)}(0,y)\equiv r_{f,0}=
\frac{1-\varepsilon_{f,0}}{1+\varepsilon_{f,0}},
\nonumber \\
&&
r_{\rm TE}^{(f,v)}(0,y)=0.
\label{eq16}
\end{eqnarray}

Expanding the logarithm in (\ref{eq15}) in power series and integrating,
one arrives at
\begin{eqnarray}
&&
{\cal F}^{(l=0)}(a,T)=-\frac{k_BT}{16\pi a^2}\sum_{n=1}^{\infty}
\frac{r_{f,0}^{2n}}{n}\int_{0}^{\infty}\!\!\!
ydye^{-ny}
\nonumber \\
&&~~~
=-\frac{k_BT}{16\pi a^2}\sum_{n=1}^{\infty}
\frac{r_{f,0}^{2n}}{n^3}=-\frac{k_BT}{16\pi a^2}{\rm Li}_3(r_{f,0}^2),
\label{eq17}
\end{eqnarray}
\noindent
where ${\rm Li}_k(z)$ is the polylogarithm function.

Under the condition (\ref{eq14}) it is easy also to estimate the contribution
of the first Matsubara frequency to the Casimir free energy of dielectric film.
For this purpose, we take into account that the TE contribution remains negligibly
small and obtain from (\ref{eq10})
\begin{eqnarray}
&&
{\cal F}^{(l=1)}(a,T)\approx\frac{k_BT}{8\pi a^2}\int_{\tau}^{\infty}\!\!\!\!\!
ydy\ln\left[\vphantom{e^{-\sqrt{y^2+(\varepsilon_{f,1}-1)\tau^2}}}
1-{r_{\rm TM}^{(f,v)}}^2(\ri\tau,y)\right.
\nonumber\\
&&~~~~~~
\times\left.
e^{-\sqrt{y^2+(\varepsilon_{f,1}-1)\tau^2}}\,\right].
\label{eq18}
\end{eqnarray}

To obtain an order of magnitude estimation, it is safe to replace $\varepsilon_{f,1}$
with $\varepsilon_{f,0}$ and $r_{\rm TM}^{(f,v)}$ with $r_{f,0}$. Then, (\ref{eq18})
results in
\begin{equation}
{\cal F}^{(l=1)}(a,T)\sim\frac{k_BT}{8\pi a^2}\int_{\tau}^{\infty}\!\!\!\!\!
ydy\ln\left(1-r_{f,0}^2e^{-\sqrt{y^2+(\varepsilon_{f,0}-1)\tau^2}}\right).
\label{eq19}
\end{equation}
\noindent
Under the condition (\ref{eq14}) one can preserve only the first term in the series
expansion of the logarithm and find
\begin{eqnarray}
&&
{\cal F}^{(l=1)}(a,T)\sim -\frac{k_BT}{8\pi a^2}r_{f,0}^2
\int_{\tau}^{\infty}\!\!\!\!\!
ydye^{-\sqrt{y^2+(\varepsilon_{f,0}-1)\tau^2}}
\nonumber \\
&&~~~~
=-\frac{k_BT}{8\pi a^2}r_{f,0}^2(1+\sqrt{\varepsilon_{f,0}}\tau)
e^{-\sqrt{\varepsilon_{f,0}}\tau}
\nonumber \\
&&~~~~
\approx-\frac{k_BT}{8\pi a^2}r_{f,0}^2\sqrt{\varepsilon_{f,0}}\tau
e^{-\sqrt{\varepsilon_{f,0}}\tau}.
\label{eq20}
\end{eqnarray}
\noindent
In the last approximate equality we have taken into account that due to the
condition (\ref{eq14}) and definition of $\tau$ (\ref{eq9}) it holds $\tau\gg 1$.
Note that the quantity (\ref{eq20}) is not classical because $\tau$ depends on $\hbar$.

Repeating similar estimations for the well known case of two thick parallel plates
separated with a vacuum gap, one arrives at the familiar result
\begin{equation}
{\cal F}_{pp}^{(l=1)}(a,T)\sim -\frac{k_BT}{8\pi a^2}r_{p,0}^2\tau e^{-\tau}
\label{eq21}
\end{equation}
\noindent
for the contribution of the first Matsubara frequency and to the same result,
as in (\ref{eq17}), for the zero-frequency contribution.

We are now in a position to compare two minimum values of $\tau$ such that
the classical limit is already reached for a pair of plates separated with
a vacuum gap and for a free-standing dielectric film made of the same material.
Let for two plates one can neglect by the contribution of the first Matsubara
frequency (\ref{eq21}) for all $\tau$ exceeding $\tau_{pp}$ (at $T=300\,$K this
is the case at $a\geqslant 5\,\mu$m, i.e., $\tau_{pp}\approx 9.22$).
Then, by comparing (\ref{eq20}) and   (\ref{eq21}), we conclude that
for a dielectric film the contribution of the first Matsubara frequency becomes
negligibly small, as compared to the zero-frequency contribution (\ref{eq17}),
at some $\tau_f$ satisfying the condition
\begin{equation}
\sqrt{\varepsilon_{f,0}}\tau_f e^{-\sqrt{\varepsilon_{f,0}}\tau_f}=
\tau_{pp} e^{-\tau_{pp}},
\label{eq22}
\end{equation}
\noindent
i.e., at $\tau_f=\tau_{pp}/\sqrt{\varepsilon_{f,0}}$. Therefore, for a dielectric
film at $T=300\,$K the classical limit is reached for rather thin films of
$a_f\approx 5/\sqrt{\varepsilon_{f,0}}\,\mu$m thickness. Numerical computations below
demonstrate a very good agreement with this result.

In the above it was assumed that the static dielectric permittivity of film
material is finite. It is common knowledge, however, that at any nonzero temperature
all dielectric materials possess some nonzero conductivity. With account of this
conductivity, the dielectric permittivity of film material takes the form \cite{2,56}
\begin{equation}
\tilde{\varepsilon}_{f}(\omega)=\varepsilon_{f}(\omega)+
\ri\frac{4\pi\sigma_0(T)}{\omega},
\label{eq23}
\end{equation}
\noindent
where $\sigma_0$ is the static conductivity which goes to zero exponentially fast
with vanishing temperature.

Substituting (\ref{eq23}) in (\ref{eq10}), for the zero-frequency contribution
to the Casimir free energy of a free-standing dielectric film one obtains
\begin{eqnarray}
&&
{\cal F}^{(l=0)}(a,T)=\frac{k_BT}{16\pi a^2}\int_{0}^{\infty}\!\!\!
ydy\ln(1-e^{-y})
\nonumber \\
&&~~~~~~~
=-\frac{k_BT}{16\pi a^2}\,\zeta(3),
\label{eq24}
\end{eqnarray}
\noindent
where $\zeta(z)$ is the Riemann zeta function. This is the same result which is known for
two plates separated by a gap, irrespective of whether they are made of a metal
described by the Drude model or dielectric with taken into account dc conductivity.
In doing so, the estimations of the role of first Matsubara frequency  in (\ref{eq20})
and (\ref{eq21}) remain unchanged.

One can conclude that the Casimir free energy of a dielectric film reaches the
classical limit,
no matter whether the dc conductivity is included in calculations or not.
If the dc conductivity is omitted, the classical limit (\ref{eq17}) is different
for different dielectrics. If the dc conductivity is included in calculations,
the classical limit (\ref{eq24}) does not depend on a specific material.
In all cases, however, the classical limit is reached for thinner films than
for two plates separated by a gap of the same width as the film thickness.
We note also that in the case of metallic films described by the plasma model
the Casimir free energy does not reach the classical limit at any film thickness
and temperature \cite{25,26,27}.

\section{Nernst heat theorem  for a free-standing dielectric film}

Taking into account that the inclusion and neglect of the dc conductivity lead to
significantly different predictions for the Casimir free energy of dielectric films, here
we investigate the thermodynamic properties of both theoretical approaches.
The point of our interest is the analytic behavior of the Casimir free energy and
entropy of a dielectric film at low temperature.

\subsection{Perfect dielectric}

It is convenient to present the Casimir free energy of a film (\ref{eq10}) as a sum
of the zero-temperature contribution and the thermal correction to it \cite{2,3}
\begin{equation}
{\cal F}(a,T)=E(a)+\Delta_T{\cal F}(a,T).
\label{eq25}
\end{equation}
\noindent
Here, the Casimir energy of a film at $T=0$ is
\begin{equation}
E(a)=\frac{\hbar c}{32\pi^2a^3}\int_{0}^{\infty}\!\!\!d\zeta\,\Phi(\zeta),
\label{eq26}
\end{equation}
\noindent
where
\begin{eqnarray}
&&
\Phi(x)=\int_{x}^{\infty}\!\!\!dy\,f(x,y),
\label{eq27} \\
&&
f(x,y)=y\left\{\ln\left[1-{r_{\rm TM}^{(f,v)}}^2(\ri x,y)
e^{-\sqrt{y^2+(\varepsilon_f(\ri x)-1)x^2}}\right]\right.
\nonumber \\
&&~~~~~
+\left.\ln\left[1-{r_{\rm TE}^{(f,v)}}^2(\ri x,y)
e^{-\sqrt{y^2+(\varepsilon_f(\ri x)-1)x^2}}\right]\right\},
\nonumber
\end{eqnarray}
\noindent
and $\Delta_T{\cal F}$ is given by
\begin{equation}
\Delta_T{\cal F}(a,T)=\frac{i\hbar c\tau}{32\pi^2a^3}\int_{0}^{\infty}\!\!\!dt
\frac{\Phi(\ri\tau t)-\Phi(-\ri\tau t)}{e^{2\pi t}-1}.
\label{eq28}
\end{equation}

The dielectric permittivity of perfect (i.e., having zero conductivity) dielectric
along the imaginary frequency axis can be represented in an oscillator form
\cite{1}
\begin{equation}
\varepsilon_f(\ri\zeta)=1+\sum_{j=1}^{N}
\frac{g_j}{\omega_j^2+\omega_c^2\zeta^2+\gamma_j\omega_c\zeta},
\label{eq29}
\end{equation}
\noindent
where $N$ is the number of oscillators, $g_j$ are the oscillator strengths,
$\omega_j$ are the oscillator frequencies, and $\gamma_j$ are the relaxation
parameters. This leads to the following permittivity at zero frequency:
\begin{equation}
\varepsilon_{f,0}=1+\sum_{j=1}^{N}
\frac{g_j}{\omega_j^2}.
\label{eq30}
\end{equation}

We are looking for the behavior of the temperature correction (\ref{eq28})
at low $T$, i.e., at $\tau\ll 1$. To find it, we take into account that the
dominant contribution to the integral in (\ref{eq28}) is given by
$t\sim 1/(2\pi)$ and, thus, $\tau t\ll 1$.
Now we determine the analytic behavior of the difference
\begin{equation}
\Delta\Phi(\tau t)=\Phi(\ri\tau t)-\Phi(-\ri\tau t)
\label{eq31}
\end{equation}
\noindent
under this condition.

As the first step, we expand the dielectric permittivity $\varepsilon_f(\ri x)$
appearing in (\ref{eq27}) up to the second power in small $x$.
Using (\ref{eq29}), one obtains
\begin{equation}
\varepsilon_f(\ri x)\approx \varepsilon_{f,0}-c_1x+c_2x^2,
\label{eq32}
\end{equation}
\noindent
where
\begin{equation}
c_1=\sum_{j=1}^{N}\frac{g_j\gamma_j\omega_c}{\omega_j^4},
\quad
c_2=\sum_{j=1}^{N}\frac{g_j\omega_c^2}{\omega_j^4}\left(1+
\frac{\gamma_j^2}{\omega_j^2}\right).
\label{eq33}
\end{equation}

Then, we perform similar expansions for the second powers of reflection
coefficients (\ref{eq11}), where $\zeta_l$ is replaced with a continuous
variable $x$ in accordance to (\ref{eq27}). For the TM mode the result is
\begin{eqnarray}
&&
{r_{\rm TM}^{(f,v)}}^2(\ri x,y)\approx
r_{f,0}^2+\frac{4r_{f,0}c_1}{(1+\varepsilon_{f,0})^2}x
-\frac{2\varepsilon_{f,0}r_{f,0}^2}{1+\varepsilon_{f,0}}\,\frac{x^2}{y^2}
\nonumber \\
&&~~~~
+\left[\frac{4c_1^2(2-\varepsilon_{f,0})}{(1+\varepsilon_{f,0})^4}-
\frac{4r_{f,0}c_2}{(1+\varepsilon_{f,0})^2}\right]x^2.
\label{eq34}
\end{eqnarray}
\noindent
Note that the perturbation expansion of ${r_{\rm TE}^{(f,v)}}^2$ starts from the
fourth order term and, thus, the TE mode does not contribute to the leading orders
considered here.

For the exponential factor, entering  (\ref{eq27}), one finds
\begin{equation}
e^{-\sqrt{y^2+(\varepsilon_{f}(ix)-1)x^2}}\approx e^{-y}\left(1-
\frac{\varepsilon_{f,0}-1}{2}\,\frac{x^2}{y}\right).
\label{eq35}
\end{equation}
\noindent
Substituting this in (\ref{eq27}) and expanding the logarithm up to the second
power in $x$, we arrive at
\begin{eqnarray}
&&
f(x,y)\approx y\ln(1-r_{f,0}^2e^{-y})-
\frac{4r_{f,0}c_1}{(1+\varepsilon_{f,0})^2}\,\frac{y x}{e^y-r_{f,0}^2}
\nonumber \\
&&
+\frac{2\varepsilon_{f,0}r_{f,0}^2}{(1+\varepsilon_{f,0})(e^y-r_{f,0}^2)}\,
\frac{x^2}{y}+\frac{\varepsilon_{f,0}-1}{2}\,\frac{r_{f,0}^2x^2}{e^y-r_{f,0}^2}
\label{eq36} \\
&&
-\left[\frac{4c_1^2(2-\varepsilon_{f,0})+2c_1^2r_{f,0}^2}{(1+\varepsilon_{f,0})^4}-
\frac{4r_{f,0}c_2}{(1+\varepsilon_{f,0})^2}\right]\,
\frac{y x^2}{e^y-r_{f,0}^2}.
\nonumber
\end{eqnarray}

Now we substitute (\ref{eq36}) in (\ref{eq27}) and perform integration with
respect to $y$. It is easily seen \cite{45} that the first, fourth and fifth terms
on the right-hand side of (\ref{eq36}) do not contribute up to the second order
in $\tau t$ inclusive in the difference (\ref{eq31}). Because of this, we focus our
attention on the contributions of the second and third terms. Considering the
second term, one finds the following contribution to the function $\Phi$ defined
in (\ref{eq27}):
\begin{eqnarray}
&&
\Phi_2(x)=-\frac{4r_{f,0}c_1x}{(1+\varepsilon_{f,0})^2}
\int_{x}^{\infty}\frac{y dy}{e^y-r_{f,0}^2}
\nonumber \\
&&~~~~
= \frac{4c_1x}{\varepsilon_{f,0}^2-1}\sum_{n=1}^{\infty}r_{f,0}^{2n}
\int_{x}^{\infty}\!\!\!\!y\,dye^{-ny}
\nonumber \\
&&~~~~
= \frac{4c_1x}{\varepsilon_{f,0}^2-1}{\rm Li}_2(r_{f,0}^2)+O(x^3).
\label{eq37}
\end{eqnarray}

This results in respective contribution to the difference (\ref{eq31})
\begin{equation}
\Delta\Phi_2(\tau t)=i\tau t
\frac{8c_1}{\varepsilon_{f,0}^2-1}{\rm Li}_2(r_{f,0}^2)+O(\tau^3t^3).
\label{eq38}
\end{equation}

In a similar way, the third term on the right-hand side of (\ref{eq36})
leads to
\begin{eqnarray}
&&
\Phi_3(x)=\frac{2\varepsilon_{f,0}r_{f,0}^2x^2}{1+\varepsilon_{f,0}}
\int_{x}^{\infty}\frac{dy}{y(e^y-r_{f,0}^2)}
\nonumber \\
&&~~~~
= \frac{2\varepsilon_{f,0}x^2}{1+\varepsilon_{f,0}}\sum_{n=1}^{\infty}r_{f,0}^{2n}
\int_{x}^{\infty}\!\!\frac{dy}{y}e^{-ny}
\nonumber \\
&&~~~~
= \frac{2\varepsilon_{f,0}x^2}{1+\varepsilon_{f,0}}\sum_{n=1}^{\infty}r_{f,0}^{2n}
{\rm Ei}(-n x),
\label{eq39}
\end{eqnarray}
\noindent
where ${\rm Ei}(z)$ is the integral exponent.

The respective contribution to the difference (\ref{eq31}) is
\begin{eqnarray}
&&
\Delta\Phi_3(\tau t)=\tau^2t^2\frac{2\varepsilon_{f,0}}{1+\varepsilon_{f,0}}
\sum_{n=1}^{\infty}r_{f,0}^{2n}i\pi
\nonumber \\
&&~~~~
=i\pi\tau^2t^2r_{f,0}^{2}\frac{1+\varepsilon_{f,0}}{2}.
\label{eq40}
\end{eqnarray}
\noindent
By summing up (\ref{eq38}) and (\ref{eq40}), we finally obtain
\begin{equation}
\Delta\Phi(\tau t)=\ri\tau t
\frac{8c_1}{\varepsilon_{f,0}^2-1}{\rm Li}_2(r_{f,0}^2)+
\ri\pi\tau^2t^2r_{f,0}^{2}\frac{1+\varepsilon_{f,0}}{2}.
\label{eq41}
\end{equation}
\noindent
This expression coincides with the result presented with no derivation in \cite{57}
for the case of two parallel dielectric plates separated with a vacuum gap.

Substituting (\ref{eq41}) in (\ref{eq28}) and integrating, one finds the
low-temperature behavior of the Casimir free energy of a dielectric film
\begin{eqnarray}
&&
\Delta_T{\cal F}(a,T)=-\frac{(k_BT)^2}{\hbar a^2}\,
\frac{{\rm Li}_2(r_{f,0}^2)}{12(\varepsilon_{f,0}^2-1)}\sum_{j=1}^{N}
\frac{g_j\gamma_j}{\omega_j^4}
\nonumber \\
&&~~~~~~~
-\frac{(k_BT)^3}{(\hbar c)^2}\,
\frac{\zeta(3)r_{f,0}^2(\varepsilon_{f,0}+1)}{4\pi}.
\label{eq42}
\end{eqnarray}

The respective Casimir entropy of a dielectric film is given by
\begin{eqnarray}
&&
S(a,T)=-\frac{\partial\Delta_T{\cal F}(a,T)}{\partial T}=
\frac{k_B^2T}{\hbar a^2}\,
\frac{{\rm Li}_2(r_{f,0}^2)}{6(\varepsilon_{f,0}^2-1)}\sum_{j=1}^{N}
\frac{g_j\gamma_j}{\omega_j^4}
\nonumber \\
&&~~~~~~~
+k_B\left(\frac{k_BT}{\hbar c}\right)^2\,
\frac{3\zeta(3)r_{f,0}^2(\varepsilon_{f,0}+1)}{4\pi}.
\label{eq43}
\end{eqnarray}

As is seen from (\ref{eq43}), the Casimir entropy of a film goes to zero
with vanishing temperature, i.e., the Nernst heat theorem is satisfied.
This makes the calculation approach with omitted dc conductivity
thermodynamically consistent.

\subsection{Account of dc conductivity}

Now we consider the dielectric film described by the dielectric permittivity
(\ref{eq23}) taking into account the dc conductivity. The Casimir free energy
in this case can be presented in the form
\begin{eqnarray}
&&
\tilde{\cal F}(a,T)={\cal F}(a,T)+\frac{k_BT}{16\pi a^2}\left[
\int_{0}^{\infty}\!\!\!y\,dy\ln(1-e^{-y})\right.
\nonumber \\
&&~~~~
\left.-\int_{0}^{\infty}\!\!\!y\,dy\ln(1-r_{f,0}^2e^{-y})\right]
+R(a,T).
\label{eq44}
\end{eqnarray}
\noindent
Here, ${\cal F}$ is the Casimir free energy (\ref{eq25}) calculated with omitted dc
conductivity  and the second term is the difference of the zero-frequency terms of
the Lifshitz formula with included and  omitted dc conductivity. The last term
$R(a,T)$ is given by
\begin{eqnarray}
&&
R(a,T)=\frac{k_BT}{8\pi a^2}\sum_{l=1}^{\infty}
\int_{\zeta_l}^{\infty}\!\!\!y\,dy
\nonumber \\
&&~~~\times
\sum_{\alpha}\left[
\ln\left(1-{{\tilde{r}_{\alpha}^{(f,v)}}}
\vphantom{\left\{\tilde{r}_{\alpha}^{(f,v)}\right\}}^2
e^{-\sqrt{y^2+(\tilde{\varepsilon}_{f,l}-1)\zeta_l^2}}\right)\right.
\nonumber \\
&&~~~~~~
\left.-
\ln\left(1-{r_{\alpha}^{(f,v)}}^2
e^{-\sqrt{y^2+(\tilde{\varepsilon}_{f,l}-1)\zeta_l^2}}\right)\right].
\label{eq45}
\end{eqnarray}
\noindent
It represents the difference of all nonzero-frequency Matsubara terms computed
with included and omitted dc conductivity.

The dielectric permittivity (\ref{eq23}) at the imaginary Matsubara frequencies
can be rewritten in the form
\begin{equation}
\tilde{\varepsilon}_{f,l}={\varepsilon}_{f,l}+
\frac{\beta(T)}{l},
\label{eq46}
\end{equation}
\noindent
where $\beta(T)=2\hbar\sigma_0(T)/(k_BT)$. As mentioned in Sec.~II,
$\sigma_0\sim\exp(-b/T)$, where $b$ is some constant
depending on the type of dielectric
material. Thus, $\beta$ goes to zero exponentially fast with decreasing
temperature. As an example, at room temperature for SiO${}_2$ it holds
$\beta(T=300\,{\rm K})\sim 10^{-12}$ \cite{58}. Because of this, the inclusion
of an additional term $\beta(T)/l$ in (\ref{eq46}) may
be considered as superfluous.
Below it is shown, however, that in the Lifshitz theory the presence of this
term, in spite of its smallness,
leads to important qualitative and quantitative consequences.

Following the same lines, as in \cite{45} (Appendix C), it can be shown
that
\begin{equation}
R(a,T)\sim\ln{T}\,e^{-b/T}
\label{eq47}
\end{equation}
\noindent
at low $T$, i.e., $R$ goes to zero together with its derivative with respect
to $T$ when $T\to 0$. Thus, performing the integrations in (\ref{eq44})
and dropping the negligibly small terms, one obtains
\begin{equation}
\tilde{\cal F}(a,T)={\cal F}(a,T)-\frac{k_BT}{16\pi a^2}\left[\zeta(3)-
{\rm Li}_3(r_{f,0}^2)\right].
\label{eq48}
\end{equation}

Calculating the negative derivative with respect to $T$ on both sides of (\ref{eq48})
and considering the limiting case of zero temperature, we arrive at
\begin{equation}
\tilde{S}(a,0)=\frac{k_B}{16\pi a^2}\left[\zeta(3)-
{\rm Li}_3(r_{f,0}^2)\right]>0.
\label{eq49}
\end{equation}
\noindent
This is a nonzero quantity depending on the parameters of a film
(the thickness and static
dielectric permittivity. This means that the Nernst heat theorem is violated.
One can conclude that the calculation procedure with taken into account dc
conductivity of film material is thermodynamically inconsistent.
Note that the Casimir entropy of metallic films satisfies the
Nernst heat theorem if the relaxation of free electrons is
omitted and violates it if the relaxation properties are
taken into account \cite{65a}.

\section{Dielectric films with negative and positive Casimir free energy}

In this section we present the results of numerical computations for the Casimir
free energy of fused silica (SiO${}_2$) and sapphire (Al${}_2$O${}_3$) dielectric
films either free-standing or covering a Au plate. All computations are performed
both with omitted and included dc conductivity of a film material demonstrating
large differences even for relatively thin films.

\subsection{Free-standing film}

\begin{figure*}[t]
\vspace*{-7cm}
\centerline{\hspace*{2.5cm}
\resizebox{0.9\textwidth}{!}{\%
\includegraphics{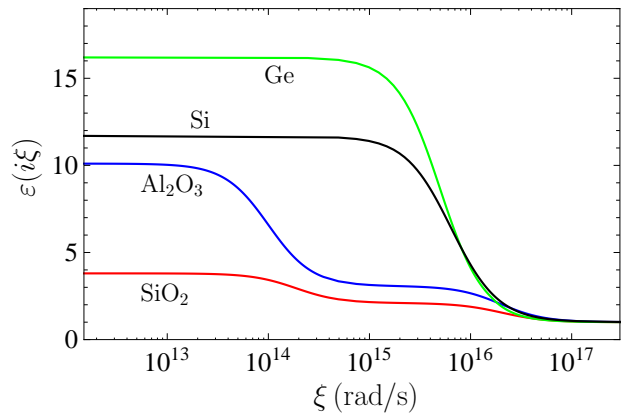}}
}
\vspace*{-6.5cm}
\caption{\label{fg1}
The dielectric permittivities of Ge, high-resistivity Si,
Al${}_2$O${}_3$ and SiO${}_2$ along the imaginary frequency
axis are shown by the four solid lines from top to bottom,
respectively.
}
\end{figure*}

The Casimir free energy of SiO${}_2$ and Al${}_2$O${}_3$ films in vacuum is computed
by (\ref{eq10}), (\ref{eq11}) and (\ref{eq8}) over the wide range of thicknesses.
For this purposes the dielectric permittivities $\varepsilon_f$ of
SiO${}_2$ and Al${}_2$O${}_3$ along the imaginary frequency axis presented in
\cite{59} have been used. They are shown by the two bottom lines in
figure~\ref{fg1}.
Specifically, for SiO${}_2$ and Al${}_2$O${}_3$ we have $\varepsilon_{f,0}=3.8$ and
10.1, respectively.

%%%%%%%%%%%%%%%%%%%%%%%%%%%%%%%%%%%%%%%%%%%%%%%%%%%%%%%%%%%%%%%%%%%%%%%
%%%%%%%__FIGURE__1__%%%%%%%%%%%%%%%%%%%%
%%%%%%%%%%%%%
%%%%%%%__FIGURE__2__%%%%%%%%%%%%%%%%%%%%
\begin{figure*}[b]
\vspace*{-8cm}
\centerline{\hspace*{2.5cm}
\resizebox{0.9\textwidth}{!}{\%
\includegraphics{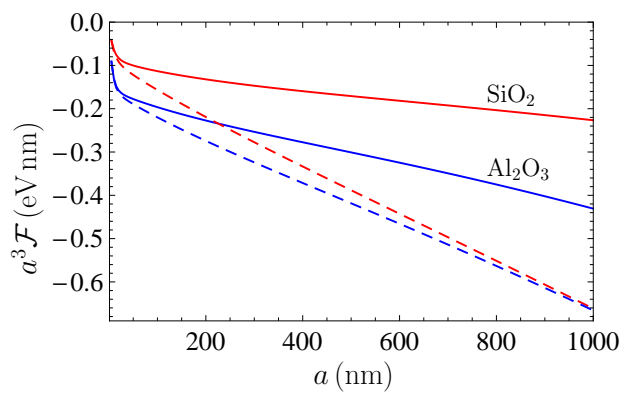}}
}
\vspace*{-6.5cm}
\caption{\label{fg2}
The Casimir free energy of free-standing films made of
SiO${}_2$ (the top pair of solid and dashed lines) and of
Al${}_2$O${}_3$ (the bottom pair of solid and dashed lines),
multiplied by the third power of film thickness, is computed
as the function of film thickness with neglected (the solid
lines) and included (the dashed lines) dc conductivity.
}
\end{figure*}
%%%%%%%%%%%%%
%%%%%%%%%%%%%%%%%%%%%%%%%%%%%%%%%%%%%%%%%%%%%%%%%%%%%%%%%%%%%%%%%%%%%%%

The computational results for the Casimir free energy per unit area of SiO${}_2$ and
 Al${}_2$O${}_3$ films multiplied by the third power of film thickness $a$ at
 $T=300\,$K are shown in figure~\ref{fg2} by the top and
 bottom solid lines, respectively,
 in the region of $a$ from 5 to 1000\,nm.
 As is seen in figure~\ref{fg2}, the larger magnitudes of the negative Casimir free energy
are obtained for a  Al${}_2$O${}_3$ film which is characterized by the larger value
of $\varepsilon$ (see figure~\ref{fg1}). According to our computational results, for
a SiO${}_2$ film the magnitude of the zero-frequency term $|{\cal F}^{(l=0)}|$
becomes approximately equal to the contribution of all nonzero Matsubara frequencies
to the Casimir free energy  $|{\cal F}^{(l\geqslant 1)}|$  for $a=465\,$nm film
thickness. The classical limit (\ref{eq17}) is reached for $a=2.6\,\mu$m, where
$|{\cal F}^{(l\geqslant 1)}|/|{\cal F}^{(l=0)}|\approx 5\times 10^{-3}$.
This is in a very good agreement with the analytic result of Sec.~II.
For a free-standing Al${}_2$O${}_3$ film
$|{\cal F}^{(l=0)}|\approx|{\cal F}^{(l\geqslant 1)}|$ for $a=340\,$nm and
$|{\cal F}^{(l\geqslant 1)}|/|{\cal F}^{(l=0)}|\approx 5\times 10^{-3}$ for
$a=2.1\,\mu$m.

The computations of the Casimir free energy of a film have been repeated using the same equations, but with the dielectric permittivity (\ref{eq46}) taking into account the dc
conductivity of the film material. Note that the results of computations do not depend
on specific values of the coefficient $\beta$ in (\ref{eq46}), but only on the fact
that it is not equal to zero. The computational results for $a^3\tilde{\cal F}$ are
shown in figure~\ref{fg2} by the top and bottom dashed lines for SiO${}_2$ and Al${}_2$O${}_3$
films, respectively.  As is seen in figure~\ref{fg2}, the Casimir free energies
computed using both approaches demonstrate considerable differences even for relatively
thin films. Thus, for  SiO${}_2$ films of $a=50$, 100, 200, 500, and 1000\,nm
thicknesses the values of $|\tilde{\cal F}|$ are greater than $|{\cal F}|$ by the factors
1.22, 1.39, 1.66, 2.27, and 2.92, respectively. For Al${}_2$O${}_3$ films the quantity
$|\tilde{\cal F}|/|{\cal F}|$ is equal to 1.07, 1.12, 1.20, 1.39, and 1.54 at the same
respective thicknesses.

As can be seen from figure~\ref{fg2}, the Casimir free energies computed for different
materials with account of the dc conductivity go to one and the same classical limit
(\ref{eq24}) with increasing film thickness. For SiO${}_2$ and Al${}_2$O${}_3$
films the classical limit is reached for $a=2.1$ and $1.9\,\mu$m, respectively, i.e.,
for somewhat thinner films than in the case of neglected dc conductivity.
The approximate equality
$|\tilde{\cal F}^{(l=0)}|\approx|{\tilde{\cal F}}^{(l\geqslant 1)}|$
holds for $a=220$ and 240\,nm for SiO${}_2$ and Al${}_2$O${}_3$
films, respectively. In the classical limit, the magnitude of the Casimir free energy
computred with included dc conductivity is substantially larger than that computed
with neglected dc conductivity. Thus,  $|\tilde{\cal F}^{(l=0)}|/|{\cal F}^{(l=0)}|=3.37$
and 1.61 for SiO${}_2$ and Al${}_2$O${}_3$ films, respectively.

\subsection{Dielectric films deposited on metallic plate}

Here, we compute the Casimir free energy of SiO${}_2$ and Al${}_2$O${}_3$
films deposited on a Au plate which can be considered as a semispace.
The computations are performed by using (\ref{eq10})--(\ref{eq12}) and (\ref{eq7}).
The dielectric permittivity of Au along the imaginary frequency axis is obtained
from the tabulated optical data \cite{56} extrapolated to lower frequencies by means
of the Drude or plasma models using the standard procedures \cite{2,3,28,29,30,31,32,33,34,35}.
In the configuration of a dielectric film deposited on a Au plate both extrapolations
lead to almost coinciding results for the Casimir free energy of a film (up to 0.3\%)
because the TE reflection coefficient at the boundary surface between a dielectric
film and vacuum vanishes at zero frequency.

%%%%%%%%%%%%%%%%%%%%%%%%%%%%%%%%%%%%%%%%%%%%%%%%%%%%%%%%%%%%%%%%%%%%%%%%%%%%
%%%%%%%__FIGURE__3__%%%%%%%%%%%%%%%%%%%%
\begin{figure}[t]
\vspace*{-1.3cm}
\centerline{\hspace*{2.5cm}
\resizebox{0.9\textwidth}{!}{\%
\includegraphics{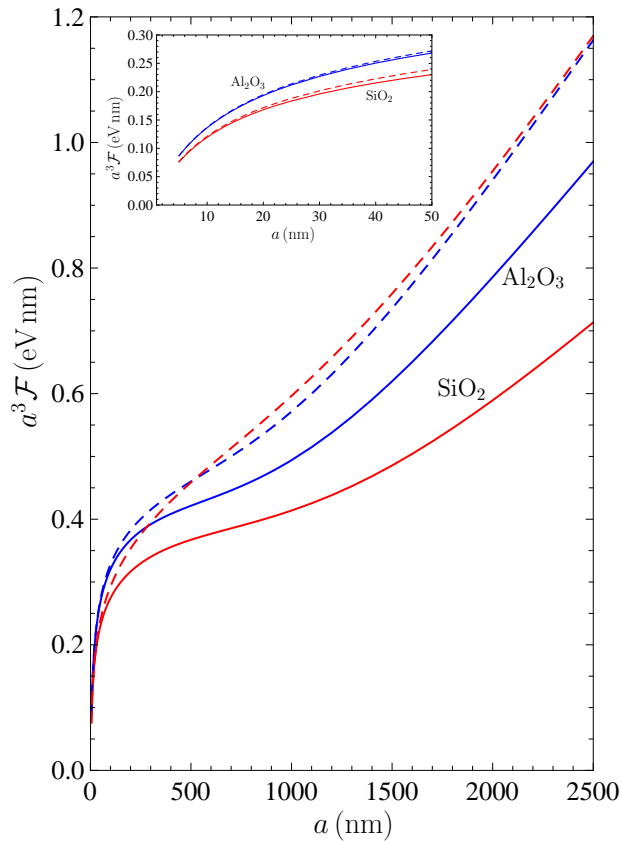}}
}
\vspace*{-6.5cm}
\caption{\label{fg3}
The Casimir free energy of SiO${}_2$ (the bottom pair of
solid and dashed lines) and of Al${}_2$O${}_3$ (the top
pair of solid and dashed lines) films deposited on a Au plate,
multiplied by the third power of film thickness, is
computed as a function of film thickness with neglected
(the solid lines) and included (the dashed lines) dc
conductivity. The case of thinnest films is shown in the
inset on an enlarged scale.
}
\end{figure}
%%%%%%%%%%%%%
%%%%%%%%%%%%%%%%%%%%%%%%%%%%%%%%%%%%%%%%%%%%%%%%%%%%%%%%%%%%%%%%%%%%%%%%%%%%
The  Casimir free energy of SiO${}_2$ and Al${}_2$O${}_3$
films deposited on a Au plate was computed with neglected dc conductivity and
multiplied by the third power of film thickness. The computational results are shown
in figure~\ref{fg3} by the bottom and top solid lines, respectively, in the region $a$ from
5 to 2500\,nm. Unlike the case of free-standing films, the Casimir free energy of
films deposited on a Au plate is positive. The approximate equality
${\cal F}^{(l=0)}\approx{\cal F}^{(l\geqslant 1)}$ holds for the film thicknesses
$a=680\,$nm (SiO${}_2$) and 550\,nm (Al${}_2$O${}_3$). The classical limit is reached
for $a=2.95\,\mu$m (SiO${}_2$ film) and $2.25\,\mu$m (Al${}_2$O${}_3$ film).
In doing so it holds
${\cal F}^{(l\geqslant 1)}/{\cal F}^{(l=0)}\approx 5\times 10^{-3}$.

In the same figure the Casimir free energy of SiO${}_2$ and Al${}_2$O${}_3$
films computed with included dc conductivity is shown for the thinnest films by the
lower and upper dashed lines, respectively. For film thicknesses $a>500\,$nm
the dashed lines change places and the one for SiO${}_2$  takes an upper position.
For both SiO${}_2$ and Al${}_2$O${}_3$
films the approximate equation
$\tilde{\cal F}^{(l=0)}\approx\tilde{\cal F}^{(l\geqslant 1)}$ holds for $a=490\,$nm.
For $a=500$ and 2000\,nm the ratio $\tilde{\cal F}/{\cal F}$ takes the values 1.25
and 1.62, respectively for SiO${}_2$ films, and 1.09 and 1.20, respectively, for
 Al${}_2$O${}_3$ films. In configurations with taken into account dc conductivity of
film material, the classical limit is reached
for $a=2.75\,\mu$m (SiO${}_2$) and $2.2\,\mu$m (Al${}_2$O${}_3$).
In the inset in figure~\ref{fg3}, the case of film thicknesses from 5 to 50\,nm is shown
in an enlarged scale for better visualization.

In the end of this section we note that the computational results for the Casimir free
energy of dielectric films obtained with taken into account dc conductivity (the
dashed lines in figures~\ref{fg2} and \ref{fg3}) are not realistic because this
calculation approach is thermodynamically inconsistent (see section~3).
Specifically, the intersection of the dashed lines in figure~\ref{fg3} is nonphysical.
It is our opinion that the physically correct values of the Casimir free energy of
dielectric films, both free-standing and deposited on metallic plate, are given
by the solid lines in figures~\ref{fg2} and \ref{fg3}.

\section{Control of the Casimir free energy of films deposited on a dielectric plate}

In this section, we calculate the Casimir free energy of SiO${}_2$, Al${}_2$O${}_3$
and Ge films deposited on the plates made of high-resistivity Si. We investigate how
the  relationship between  ${\cal F}$ and $a$
depends on the thickness of a Si plate.
It is shown that the values of the free energy and even its sign depend critically on
the thickness of a dielectric substrate. We demonstrate also that one can control the
Casimir free energy of a film by illuminating the plate with laser pulses.

\subsection{Silica films}

We have performed numerical computations of the Casimir free energy of SiO${}_2$ film
of thickness $a$ deposited on the plate of finite thickness $d$ made of high-resistivity
Si with $\varepsilon_{p,0}=11.7$. For this purpose (\ref{eq10}) has been used
where the reflection coefficients are given by (\ref{eq11}) and (\ref{eq4}).
The latter equation was expressed in terms of the dimensionless variables with account
of (\ref{eq13}). We show that if the plate is made of dielectric material its
thickness may have a profound effect on the Casimir free energy of a film.

The dielectric permittivity of SiO${}_2$ was already used above in computations
of section~4. The dielectric permittivity of high-resistivity Si along the imaginary
frequency axis is obtained by means of the Kramers-Kronig relation from the
tabulated optical data Si \cite{56a}. It is shown in figure~\ref{fg1} as a function of
frequency. In this section, we do not include the contribution of dc conductivity
in the permittivities of dielectric materials because this inclusion leads to
contradictions with thermodynamics (see section~3).

\begin{figure}[b]
\vspace*{-7.5cm}
\centerline{\hspace*{2.5cm}
\resizebox{0.9\textwidth}{!}{\%
\includegraphics{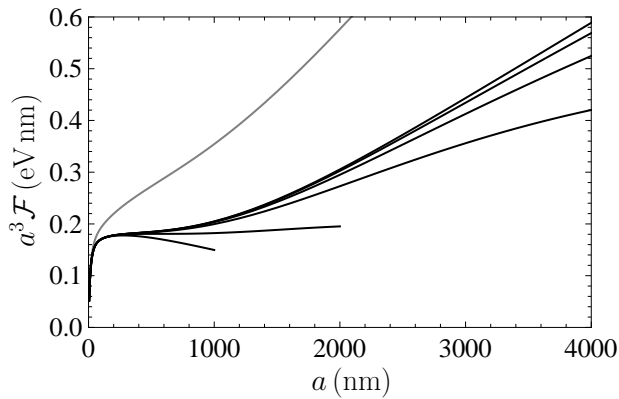}}
}
\vspace*{-6.5cm}
\caption{\label{fg4}
The Casimir free energy of SiO${}_2$ films deposited on Si
plates of different thicknesses equal to 1, 2, 5, 10, 20, and
$\geqslant 100\,\mu$m, multiplied by the third power of film
thickness, is shown by the six lines counted from the figure
bottom, respectively. The top line shows the Casimir free
energy of SiO${}_2$ film when the Si plate is illuminated
with laser pulses.
}
\end{figure}

In figure~\ref{fg4} we present the computational results for the free energy per unit
area of SiO${}_2$ film deposited on Si plate of finite thickness $d$ as a  function
of  film thickness $a$. These results are multiplied by the third power of the film
thickness. The six lines counted from the figure bottom are plotted for the plates
of $d=1$, 2, 5, 10, 20, and $\geqslant 100\,\mu$m thickness. In so doing, it is assumed
that $a < d$, i.e., that the plate is thicker than a film. The top line in
figure~\ref{fg4} demonstrates the computational results for the case of Si plate
illuminated with laser pulses (see below).
%%%%%%%%%%%%%%%%%%%%%%%%%%%%%%%%%%%%%%%%%%%%%%%%%%%%%%%%%%%%%%%%%%%%%%%%%%%%%
%%%%%%%__FIGURE__4__%%%%%%%%%%%%%%%%%%%%
%%%%%%%%%%%%%
%%%%%%%%%%%%%%%%%%%%%%%%%%%%%%%%%%%%%%%%%%%%%%%%%%%%%%%%%%%%%%%%%%%%%%%%%%%%%

As is seen in figure~\ref{fg4}, the Casimir free energy of SiO${}_2$ film  is positive,
but decreases with decreasing thickness of Si plate. This can be explained by the
following. Computations show that the contribution of all nonzero Matsubara
frequencies, ${\cal F}^{(l\geqslant 1)}$, into the total Casimir free energy is
almost independent on the thickness of a Si plate $d$. Thus,
for $a=1\,\mu$m ${\cal F}^{(l\geqslant 1)}$
changes for only 0.2\% when $d$ increases from $1\,\mu$m to $\infty$.
One concludes that the dependence of ${\cal F}$ on $d$ is mostly determined by the
zero-frequency contribution ${\cal F}^{(l=0)}$. The latter depends on $d$
through the reflection coefficient $R_{\rm TM}^{(f,p)}(0,y)$ defined in
(\ref{eq4}) and (\ref{eq13}) as
\begin{equation}
R_{\rm TM}^{(f,p)}(0,y)=\frac{r_{\rm TM}^{(f,s)}(0)+
r_{\rm TM}^{(p,v)}(0)\,e^{-dy/a}}{1+r_{\rm TM}^{(f,s)}(0)
r_{\rm TM}^{(p,v)}(0)\,e^{-dy/a}}.
\label{eq50}
\end{equation}

For the film and plate materials under consideration, from (\ref{eq11}) and
(\ref{eq12}) we have
\begin{eqnarray}
&&
r_{\rm TM}^{(f,s)}(0)=\frac{\varepsilon_{p,0}-
\varepsilon_{f,0}}{\varepsilon_{p,0}+\varepsilon_{f,0}}
\approx 0.51,
\nonumber \\
&&
r_{\rm TM}^{(p,v)}(0)=\frac{1-
\varepsilon_{p,0}}{1+\varepsilon_{p,0}}
\approx -0.84.
\label{eq51}
\end{eqnarray}

Taking into account that the dominant contribution to the zero-frequency term
of the free energy (\ref{eq10}) is given by $y\sim 1$, one concludes that both the
numerator and denominator in (\ref{eq50}) are positive. For the fixed film
thickness $a$ the relative role of the negative coefficient $r_{\rm TM}^{(p,v)}(0)$
increases with decreasing $d$ and, as a result, the free energy ${\cal F}$
decreases. In doing so, it remains positive, because, according to ({\ref{eq16}),
another reflection coefficient in (\ref{eq10}) is negative,
$r_{\rm TM}^{(f,v)}(0)\approx -0.58$.

It is instructive to compute the quantity $s_0$ defined as
\begin{equation}
{\cal F}^{(l=0)}(a,T)=\frac{k_BT}{8\pi a^2}\,s_0(a,T),
\label{eq52}
\end{equation}
\noindent
which is in fact the function of the ratio $a/d$. For SiO${}_2$ film deposited on a Si
plate the computational results are shown by the top line in figure~\ref{fg5}.
As is seen in figure~\ref{fg5}, the quantity $s_0$ is positive for all values of $a/d$
in accordance with the computational results presented in figure~\ref{fg4}.
Below we show that this quantity determines the qualitative behavior of the
Casimir free energy of dielectric films deposited on a dielectric plate.
%%%%%%%%%%%%%%%%%%%%%%%%%%%%%%%%%%%%%%%%%%%%%%%%%%%%%%%%%%%%%%%%%%%%%%%%%%%%%%%
%%%%%%%__FIGURE__5__%%%%%%%%%%%%%%%%%%%%
\begin{figure}[t]
\vspace*{-7.5cm}
\centerline{\hspace*{2.5cm}
\resizebox{0.9\textwidth}{!}{\%
\includegraphics{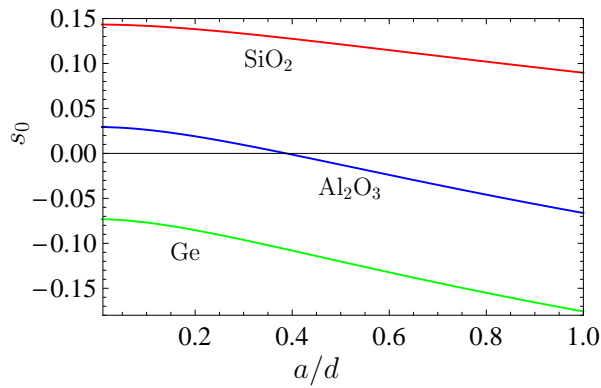}}
}
\vspace*{-6.5cm}
\caption{\label{fg5}
The normalized zero-frequency contribution to the Casimir
free energy of SiO${}_2$, Al${}_2$O${}_3$ and Ge films
deposited on Si plate is shown by the three lines from top
to bottom, respectively, as a function of the ratio of film
to plate thicknesses.
}
\end{figure}
%%%%%%%%%%%%%
%%%%%%%%%%%%%%%%%%%%%%%%%%%%%%%%%%%%%%%%%%%%%%%%%%%%%%%%%%%%%%%%%%%%%%%%%%%%%%%

Now  we show that it is possible to significantly increase the
Casimir free energy of dielectric film deposited on a plate and make it independent
on the plate thickness. For this purpose one can use the illumination by laser pulses
from a 514\,nm Ar laser incident on the bottom surface of the plate.
Note that the effect of illumination was considered in the system
Au-ethanol-Si \cite{67a}.
Here, we consider the same parameters as
in the experiment on optically modulated
Casimir forces \cite{2,3,40}. The high-resistivity Si was $p$-doped with  charge
carrier density $n_0\approx 5\times 10^{14}\,\mbox{cm}^{-3}$ \cite{56a}.
In the presence of pulse an equilibrium concentration of charge carriers
(electrons and holes) of  $n_e\approx 2\times 10^{19}\,\mbox{cm}^{-3}$ is
 rapidly established in the plate with some definite absorbed power \cite{2,3,40}.
 Then, the dielectric permittivity of the plate is given by
 \begin{equation}
 \tilde{\varepsilon}_p(\ri\zeta)=\varepsilon_p(\ri\zeta)+
 \frac{\tilde{\omega}_{p(e)}^2}{\zeta[\zeta+\tilde{\gamma}_{(e)}]}+
\frac{\tilde{\omega}_{p(p)}^2}{\zeta[\zeta+\tilde{\gamma}_{(p)}]}.
\label{eq53}
\end{equation}
\noindent
Here, $\tilde{\omega}_{p(e,p)}={\omega}_{p(e,p)}/\omega_c$ are the dimensionless
plasma
frequencies for electrons and holes, respectively, and the dimensional ones are
equal to ${\omega}_{p(e)}\approx 5.0\times 10^{14}\,$rad/s and
${\omega}_{p(p)}\approx 5.6\times 10^{14}\,$rad/s \cite{2,3,40}. In a similar way,
the dimensionless and dimensional relaxation parameters are
$\tilde{\gamma}_{(e,p)}={\gamma}_{(e,p)}/\omega_c$,
${\gamma}_{(e)}\approx 1.8\times 10^{13}\,$rad/s and
${\gamma}_{(p)}\approx 5.0\times 10^{12}\,$rad/s \cite{60}.
Note that $\tilde{\varepsilon}_p$ in (\ref{eq53}) notates the permittivity of Si
in the metallic state, and not an inclusion of
the dc conductivity in the dielectric state.

The Casimir free energy of SiO${}_2$ film in the presence of laser light
 on a Si plate was
computed by (\ref{eq10}), (\ref{eq11}), (\ref{eq4}), and (\ref{eq53}),
as a function
of film thickness for different thicknesses of Si plate. Almost the
same computational
results are obtained if one puts in (\ref{eq53}) $\tilde{\gamma}_{(e,p)}=0$,
i.e., if the plasma model in place of the Drude model is used.
The computational results are shown by the top line in figure~\ref{fg4}.
They are independent of the plate thickness. The classical limit is reached for film
thicknesses $a\geqslant 2500\,$nm.

\subsection{Sapphire films}

Now we show that the Casimir free energy of a dielectric film
deposited on Si plate can be significantly different depending
on the film material. For this purpose, we consider
Al${}_2$O${}_3$ films covering a Si plate and perform numerical
computations using the same (\ref{eq10}), ~(\ref{eq11}),
and ~(\ref{eq4}), as for SiO${}_2$ films.

The computational results for the Casimir free energy of
Al${}_2$O${}_3$ film deposited on Si plate of 1, 2, 5, 10, 20,
50, and $\geqslant 100\,\mu$m thickness are multiplied by $a^{3}$
and shown in figure~\ref{fg6}(a) by seven lines counted from the
figure bottom, respectively, as a function of film thickness
under a condition $a < d$. Similar to the case of SiO${}_2$
films, the Casimir free energy of Al${}_2$O${}_3$ films for
fixed $a$ decreases with decreasing thickness $d$ of a Si plate.
However, unlike the case of SiO${}_2$ films, the Casimir free
energy of Al${}_2$O${}_3$ films takes both positive and
negative values. This is explained by the following.
%%%%%%%%%%%%%%%%%%%%%%%%%%%%%%%%%%%%%%%%%%%%%%%%%%%%%%%%%%%%%%%%
%%%%%%%__FIGURE__6__%%%%%%%%%%%%%%%%%%%%
\begin{figure}[t]
\vspace*{-3cm}
\centerline{\hspace*{.5cm}
\resizebox{\textwidth}{!}{\%
\includegraphics{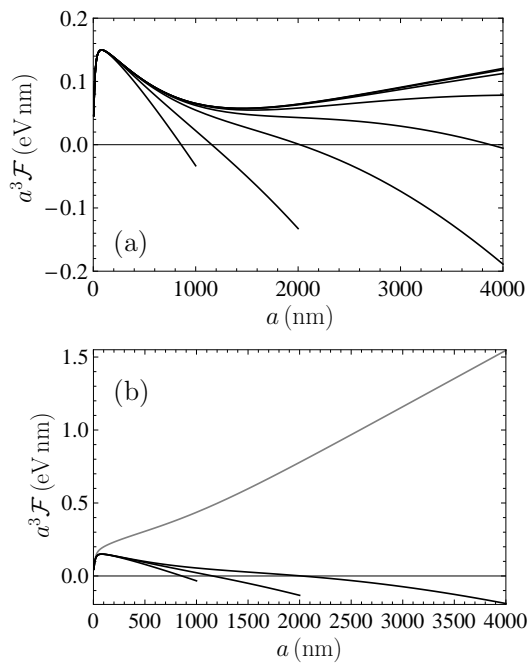}}
}
\vspace*{-10cm}
\caption{\label{fg6}
(a) The Casimir free energy of Al${}_2$O${}_3$ films deposited
on Si plates of different thicknesses equal to 1, 2, 5, 10, 20,
50, and $\geqslant 100\,\mu$m, multiplied by the third power of
film thickness, is shown by the seven lines counted from the
figure bottom, respectively. (b) The top line shows the Casimir
free energy of Al${}_2$O${}_3$ film when the Si plate is
illuminated with laser pulses. The three lines counted from
the figure bottom reproduce that ones plotted in figure~6(a).
}
\end{figure}
%%%%%%%%%%%%%
%%%%%%%%%%%%%%%%%%%%%%%%%%%%%%%%%%%%%%%%%%%%%%%%%%%%%%%%%%%%%%%%%

First we note that the dependence of ${\cal F}$ on $d$ is again
determined by the contribution of ${\cal F}^{(l=0)}$. For
Al${}_2$O${}_3$ film of $a = 1\,\mu$m thickness the contribution
of ${\cal F}^{(l\geqslant 1)}$ to ${\cal F}$ changes in the
limits of 0.3\% when $d$ varies from 1\,$\mu$m to $\infty$.
Then we take into account that the reflection coefficients in
(\ref{eq50}) are now equal to $r_{\rm TM}^{(f,s)}(0)\approx
0.073$ and  $r_{\rm TM}^{(p,v)}(0)\approx -0.84$. The smallness
of $r_{\rm TM}^{(f,s)}(0)$ results in the fact that even for
$d/a = 1$ the numerator in (\ref{eq50}) is negative
whereas the denominator is positive. Taking into account that
for Al${}_2$O${}_3$ film $r_{\rm TM}^{(f,v)}(0)\approx -0.82$, one
obtains from Eq.~(\ref{eq10}) that the Casimir free energies of
sufficiently thick films are negative. According to
figure~\ref{fg6}(a), the Casimir free energy of Al${}_2$O${}_3$
films of thicknesses equal to 855, 1150, 2050, and 3000\,nm
vanishes if these films are deposited on Si plates with
thicknesses of 1, 2, 5, and 10\,$\mu$m, respectively. Thus, by
matching the thicknesses of a film and a plate, one can control
the sign and value of the Casimir free energy. The classical
limit is again reached for film thicknesses exceeding 2500 nm.

Finally, we consider the effect of illumination of a Si plate
with laser pulses. The same parameters as above and the
dielectric permittivity (\ref{eq53}) for Si
in the presence of laser light have been
used in computations. The computational
results for the Casimir free energy of Al${}_2$O${}_3$ film
deposited on illuminated Si plates of different thicknesses
are shown in figure~\ref{fg6}(b) by the top line as a function
of film thickness. For comparison purposes, the same Casimir
free energies, as in figure~\ref{fg6}(a), are shown by the three
lines counted from the figure bottom for the films deposited
on Si plates of 1, 2, and 5 nm thickness, respectively, with
no illumination. It is seen that in the presence of laser
light the Casimir free energy of a film is positive for any
film thickness and does not depend on the thickness of a plate.

\subsection{Germanium films}

Here, we consider the Casimir free energy of Ge films deposited
on Si plates. The dielectric permittivity of Ge along the
imaginary frequency axis is taken from \cite{61}. It is
shown by the top line in figure~\ref{fg1}. The static dielectric
permittivity of Ge is equal to $\epsilon_{f}\approx 16.2$.

The computational results for the Casimir free energy of Ge
film deposited on Si plates of 1, 2, 5, 10, 20, and
$\geqslant 100\,\mu$m thickness are multiplied by $a^{3}$ and
shown in figure~\ref{fg7}(a) by six lines counted from the
figure bottom, respectively, as a function of film thickness
under a condition $a < d$. From figure~\ref{fg7}(a) it is seen
that the Casimir free energy of a film is negative and decreases
with decreasing film thickness.
%%%%%%%%%%%%%%%%%%%%%%%%%%%%%%%%%%%%%%%%%%%%%%%%%%%%%%%%%
%%%%%%%__FIGURE__7__%%%%%%%%%%%%%%%%%%%%
\begin{figure}[b]
\vspace*{-3.2cm}
\centerline{\hspace*{.5cm}
\resizebox{\textwidth}{!}{\%
\includegraphics{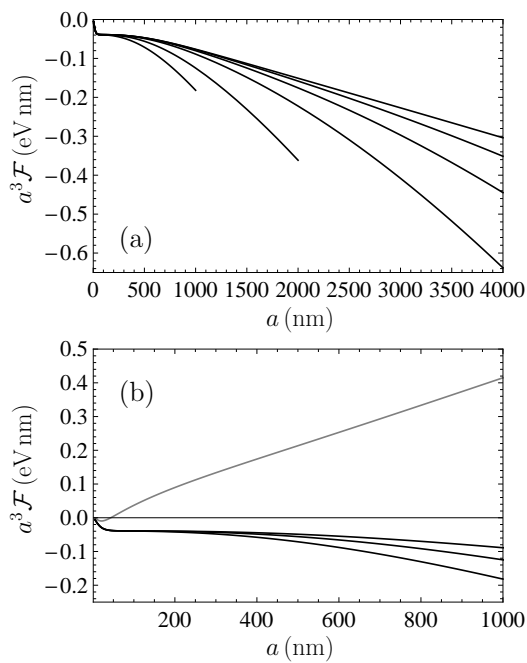}}
}
\vspace*{-10cm}
\caption{\label{fg7}
(a) The Casimir free energy of Ge films deposited
on Si plates of different thicknesses equal to 1, 2, 5, 10, 20,
and $\geqslant 100 \mu$m, multiplied by the third power of
film thickness, is shown by the six lines counted from the
figure bottom, respectively. (b) The top line shows the Casimir
free energy of Ge film when the Si plate is illuminated with
laser pulses. The three lines counted from the figure bottom
reproduce that ones plotted in figure~7(a).
}
\end{figure}
%%%%%%%%%%%%%
%%%%%%%%%%%%%%%%%%%%%%%%%%%%%%%%%%%%%%%%%%%%%%%%%%%%%%%%%

The dependence of the Casimir free energy on $d$ is again
determined by the zero-temperature contribution
${\cal F}^{(l=0)}$. Computations show that
${\cal F}^{(l\geqslant 1)}$ decreases for 1\% when the plate
thickness $d$ increases from $1\,\mu$m to $\infty$. For Ge the
reflection coefficients in (\ref{eq50}) are equal to
$r_{\rm TM}^{(f,s)}(0)\approx -0.16$ and
$r_{\rm TM}^{(p,v)}(0)\approx -0.84$, i.e., are both negative.
This leads to $R_{\rm TM}^{(f,p)}(0) < 0$. Taking into account
that for Ge $r_{\rm TM}^{(f,v)}(0)\approx -0.88$, one finds that
${\cal F}^{(l=0)} < 0$.

In the case of Ge films, illumination of a Si plate with laser
light makes the strongest effect on the Casimir free energy.
In this case the reflection coefficient $r_{\rm TM}^{(f,s)}(0)$
on the boundary plane between Ge and Si changes not only its
magnitude, but the sign as well by becoming positive in the
presence of laser light. This leads to the change of a sign of
the Casimir free energy.

In figure~\ref{fg7}(b) the computational results for the Casimir
free energy of Ge film multiplied by the third power of film
thickness in the presence of laser light on a Si plate are
shown by the top line as a function of film thickness. For
comparison purposes, the Casimir free energies of Ge films
deposited on nonilluminated Si plates of 1, 2, and 5\,$\mu$m
thickness are plotted by the three lines counted from the
figure bottom, respectively.

As is seen in figure~\ref{fg7}(b), the Casimir free energy of
the thinnest Ge films deposited on illuminated Si plates is
negative. The values of the free energy do not depend on the
thickness of a Si plate. For a film of 44\,nm thickness,
the Casimir free energy takes the zero value and for thicker
films becomes positive. For Ge films deposited on the
illuminated Si plate the classical limit is reached for the
film thickness equal to 1600\,nm.

\section{Conclusions and discussion}

In the foregoing, we have investigated the Casimir free energy
of dielectric films, both free-standing in vacuum and deposited
on material plates. It is shown that the Casimir free energy of
free-standing films made of SiO${}_2$ and Al${}_2$O${}_3$
depends considerably on whether one neglects by the role of dc
conductivity of film material or includes it in computations.
The obtained differences in theoretical predictions are much
larger than for the case of two parallel plates separated with
a gap. In both theoretical approaches the classical limit is
reached, but for thinner films, as compared to the gap width in
the configuration of two parallel plates. If the dc conductivity
is neglected, the classical limit depends on film material. With
included dc conductivity, the classical limit demonstrates a
universal behavior.

To find a convincing theoretical argument in favor of one or
other calculation approach, we have investigated analytic behaviors of
the Casimir free energy and entropy of a dielectric film at
arbitrarily low temperature. It is shown that if the dc
conductivity is neglected the Casimir entropy goes to zero with
vanishing temperature in accordance to the third law of
thermodynamics (the Nernst heat theorem). If the dc conductivity
of film material is included in calculations, the Casimir
entropy at zero temperature is shown to be equal to a positive
quantity depending on the film parameters in violation of the
Nernst heat theorem.

Similar situation takes place for two dielectric plates separated
with a vacuum gap, where the inclusion of dc conductivity results
in violation of the Nernst theorem \cite{45,46,47,48,49}. In this
respect one cannot overemphasize the fact that measurements of
the Casimir interaction in dielectric systems are in good
agreement with the thermodynamically consistent approach and
exclude that one violating the Nernst theorem \cite{40,41,42,43,44}.
The problem of consistency with thermodynamics is interesting also for
some configurations involving graphene and topological insulators
which have been used to show the possibility of Casimir repulsion
through a vacuum gap \cite{70,71,72}. Using the formalism of the
polarization tensor \cite{73,74,75,75a}, it was shown \cite{76}
that the Casimir entropy of graphene satisfies the Nernst heat
theorem. Then, it is hardly surprising that measurements of the
Casimir interaction in graphene system \cite{77} were found in
agreement with theoretical predictions exploiting the same
formalism \cite{78}. The validity of the Nernst theorem for the
surface states of topological insulators is an open question
to be solved in near future.

We have computed the Casimir free energy of SiO${}_2$ and
Al${}_2$O${}_3$ films deposited on Au plates. It was shown that
the computational results are almost independent on whether the
Drude or the plasma model is used to describe the dielectric
permittivity of Au at low frequencies. The obtained free
energies are also independent on a thickness of Au plate
provided it is larger than several tens nanometers. However, the
Casimir free energy of dielectric films deposited on a Au plate
are shown to be sensitive to an account or neglect of the dc
conductivity of film material. According to our argumentation,
the physically correct results are obtained within the calculation
approach which ignores the dc conductivity of dielectric films.

Special attention was paid to the possibility of controlling
the Casimir free energy of dielectric films made of different
materials, such as SiO${}_2$, Al${}_2$O${}_3$ and Ge. For this
purpose, dielectric films deposited on dielectric
(high-resistivity Si) plates of finite thickness were considered.
It is shown that for a fixed film thickness the Casimir free
energy of a film decreases with increasing thickness of the
plate. According to our results, the free energy of a film
deposited on a Si plate can be positive (SiO${}_2$ films),
positive, zero or negative depending on the film and plate
thicknesses (Al${}_2$O${}_3$ films) and negative (Ge films).
Illumination of the Si plate with laser light of Ar laser
results in an independent of the plate thickness Casimir free
energy of a film, which is mostly positive (with exception of
only the case of very thin Ge films).

Taking into account a tendency for further miniaturization
in various technologies, the above results may be useful in
numerous applications of dielectric films in optics and
microelectronics, where the fluctuation-induced forces are
coming to play their role.

%%%%%%%%%%%%%%%%%%%%%%%%%%%%%%%%%%%%%%%
\ack{The work of V.M.M. was partially supported by the Russian Government
Program of Competitive Growth of Kazan Federal University.}
%%%%%%%%%%%%%%%%%%%%%%%%%%%%%%%%%%

%%%%%%%%%%%%%%
\end{document}